\documentclass[conference]{IEEEtran}
\usepackage{cite}
\usepackage{amsmath,amssymb,amsfonts}
\usepackage{algorithmic}
\usepackage{graphicx}
\usepackage{textcomp}
\usepackage{xcolor}
\def\BibTeX{{\rm B\kern-.05em{\sc i\kern-.025em b}\kern-.08em
    T\kern-.1667em\lower.7ex\hbox{E}\kern-.125emX}}

\usepackage{subcaption}
\usepackage{booktabs}
\usepackage{multirow}
\begin{document}

\title{A Comprehensive Analysis of Real-World Accelerometer Data Quality in a Global Smartphone-based Seismic Network}


\author{\IEEEauthorblockN{Yawen Zhang\IEEEauthorrefmark{1},
Qingkai Kong\IEEEauthorrefmark{2}, Tao Ruan\IEEEauthorrefmark{1}, Qin Lv\IEEEauthorrefmark{1} and Richard Allen \IEEEauthorrefmark{3}}
\IEEEauthorblockA{\IEEEauthorrefmark{1}Department of Computer Science, University of Colorado Boulder, Boulder, Colorado, USA\\ \IEEEauthorrefmark{2}Lawrence Livermore National Laboratory, Livermore, California, USA\\
\IEEEauthorrefmark{3}Berkeley Seismological Laboratory, Berkeley, California, USA\\
Email: \IEEEauthorrefmark{1}\{tao.ruan, yawen.zhang, qin.lv\}@colorado.edu,
\IEEEauthorrefmark{2}kongqk@berkeley.edu, \IEEEauthorrefmark{3}rallen@berkeley.edu}}


\makeatletter
\let\@oldmaketitle\@maketitle
\renewcommand{\@maketitle}{\@oldmaketitle
  \centering
  \includegraphics[width=0.7\textwidth]
    {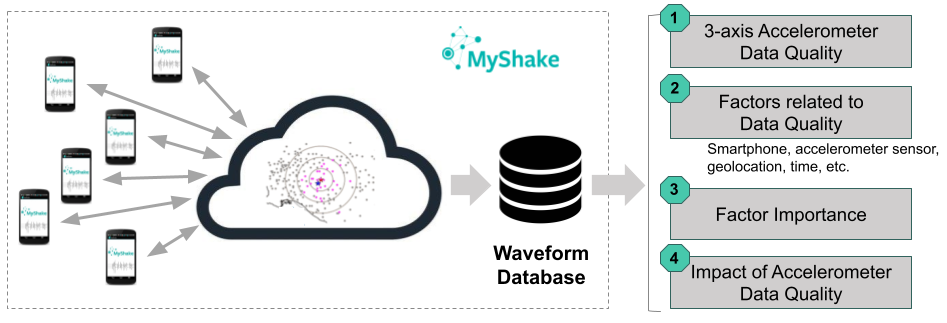}\bigskip}
  \label{fig:graphical_abstract}
\makeatother

\maketitle

\raggedright 
\frenchspacing

\begin{abstract}

The proliferation of low-cost sensors in smartphones has facilitated numerous applications; however, large-scale deployments often encounter performance issues. Sensing heterogeneity, which refers to varying data quality due to factors such as device differences and user behaviors, presents a significant challenge. In this research, we perform an extensive analysis of 3-axis accelerometer data from the MyShake system, a global seismic network utilizing smartphones. We systematically evaluate the quality of approximately 22 million 3-axis acceleration waveforms from over 81 thousand smartphone devices worldwide, using metrics that represent sampling rate and noise level. We explore a broad range of factors influencing accelerometer data quality, including smartphone and accelerometer manufacturers, phone specifications (release year, RAM, battery), geolocation, and time. Our findings indicate that multiple factors affect data quality, with accelerometer model and smartphone specifications being the most critical. Additionally, we examine the influence of data quality on earthquake parameter estimation and show that removing low-quality accelerometer data enhances the accuracy of earthquake magnitude estimation.

\end{abstract}


\section{Introduction}

Smartphones, equipped with a collection of sensors, have increasingly been employed in various applications.~\cite{tan2016riding,allouch2017roadsense,elhamshary2018crowdmeter,wang2020predicting}. In recent years, researchers have been exploring using smartphones in disaster-related applications, for example, earthquake detection~\cite{kong2016myshake,allen2019earthquake,Earle2010,Minson2015,finazzi2016earthquake,Bossu2018}. Traditionally,
the monitoring of earthquakes relies on high-quality seismometers. Seismic networks built with a good density of seismometers are a prerequisite for developing an Earthquake Early Warning (EEW) system~\cite{kong2016myshake}. It is expensive to deploy and maintain such networks in many underdeveloped regions. While smartphone-based accelerometer data are primarily used for Human Activity Recognition (HAR)~\cite{kwapisz2011activity,xu2011mems,bayat2014study,khan2018scaling}, the wide availability of low-cost accelerometers on smartphones offers novel approaches for earthquake detection~\cite{allen2012transforming}. To build a global seismic network, the MyShake system~\footnote{https://myshake.berkeley.edu/} has been developed to leverage accelerometers on smartphones to detect earthquake-like motions~\cite{kong2016myshake2,kong2016myshake}. The MyShake system was launched in 2016 by the University of California, Berkeley. To date, it has recorded over 1000 earthquakes worldwide~\cite{kong2020toward1}. The MyShake application (available in the iTunes App and Google Play stores) has been downloaded by more than 2.5 million users, and more than 500,000 phones interact with the system each day. MyShake also delivers USGS ShakeAlert earthquake early warning messages to California, Oregon, and Washington~\cite{strauss2020myshake}. Building upon MyShake, Google has recently announced the Android Earthquake Alerts System~\footnote{https://blog.google/products/android/earthquake-detection-and-alerts/}~\cite{allen2022global} that makes use of billions of Android phones globally as mini seismometers for earthquake detection. This effort would greatly expand the capability of mobile sensing for earthquake detection. MyShake has also started to deliver USGS ShakeAlert earthquake early warning messages to the state of California~\cite{allen2020}. \bigskip



Like many other mobile sensing applications, when deployed at a large scale, MyShake faces the performance challenge arising from sensing heterogeneity~\cite{kong2019earthquake}. The accelerometer data collected exhibit varying data quality due to different devices, user behaviors, etc. Previous studies have shown that such sensing heterogeneity can significantly impair the performance of accelerometer-based applications~\cite{stisen2015smart,blunck2016activity}. Currently, the MyShake system relies on seismologists to manually review all the waveforms and remove those with significant quality issues, e.g., missing data, spikes~\cite{kong2019assessing}, which is impractical in large-scale deployments. It is crucial to automate the process of quality assessment, and gain a comprehensive understanding of the primary quality concerns associated with accelerometer data. Furthermore, accelerometer data quality may be influenced by a variety of factors, including the device, accelerometer sensor, user's location, and more. Previous studies has investigated influencing factors like the device and accelerometer sensor, but with a small number of devices in the analysis (e.g., 13 smartphones from 4 manufacturers in \cite{stisen2015smart}). Real-world applications such as MyShake are dealing with a much larger number of devices (81 thousand smartphones) as well as accelerometer sensors. Our study represents the first comprehensive, large-scale analysis of accelerometer data quality. The findings would be beneficial for not only MyShake but also other accelerometer-based applications. \bigskip



To summarize, our study makes the following contributions:

\begin{itemize}
    \item We investigate real-world, large-scale 3-axis accelerometer data collected by the MyShake system, assessing their quality based on parameters such as sampling rate and noise level. 
    \item We explore an extensive range of factors, including smartphone manufacturer, accelerometer sensor, smartphone specifications (RAM, battery, etc.), geolocation, and trigger time, as to understand their relationships with accelerometer data quality. 
    \item We assess the importance of various factors by employing them to predict accelerometer data quality. The findings suggest that the quality of accelerometer data is influenced by multiple factors, with the accelerometer model and smartphone specifications being the most important ones. 
    \item We examine the effect of accelerometer data quality on earthquake parameter estimation by applying quality control in real-world earthquake events. The results demonstrate that by filtering out poor-quality accelerometer data, the accuracy of earthquake magnitude estimations can be improved. 
\end{itemize}

\section{Related Work}

\subsection{New Approaches for Earthquake Detection}

Traditionally, earthquake detection relies on high-quality seismometers. In recent years, there have been many studies that explore new ways to detect earthquakes. Sakaki \textit{et al.}~\cite{sakaki2010earthquake,sakaki2012tweet} and Earle \textit{et al.}~\cite{Earle2010} leverage Twitter data to estimate the locations of earthquake events, and build an earthquake reporting system upon that. Avvenuti \textit{et al.}~\cite{avvenuti2014ears} make use of real-time Twitter messages to detect earthquake events, and mine the message content to discover knowledge about the consequences of those events. More recent studies explore the potential of utilizing low-cost accelerometers, which are found in devices such as smartphones and connected vehicles, for the purpose of earthquake detection~\cite{dashti2014evaluating,finazzi2016earthquake,kong2016myshake,Minson2015,geotab}. MyShake is such an application that leverages these low-cost accelerometers in smartphones as seismometers to detect earthquake events.




\subsection{Accelerometer Data Quality}

When deployed in the real world, mobile sensing usually face the challenge arises from varying data quality. In the HAR application, Stisen \textit{et al.}~\cite{stisen2015smart} analyze several types of sensing heterogeneities like sensor bias, sampling rate, and sampling rate instability. Martinez \textit{et al.}~\cite{martinez2020quality} examine wearable data quality by looking into metrics related to data gaps, replacements, and wearable energy levels. A variety of sources can contribute to data quality variability, including the device, sensor, operating system (OS), user behavior, etc. Although previous studies have explored these sources, the scale of examination has been limited. For instance, Stisen \textit{et al.}~\cite{stisen2015smart} investigate heterogeneity sources such as accelerometer sensor and CPU load across 13 phone models from four manufacturers. Min \textit{et al.}~\cite{min2019closer} analyze device and temporal variability in a multi-device setting with an accelerometer dataset from 15 participants performing seven activities. Distinguished from previous studies, our study presents the first comprehensive, large-scale analysis of accelerometer data quality. Additionally, we thoroughly examine an extensive range of factors that could potentially contribute to data quality variability.



\subsection{Quality-Aware Mobile Crowdsensing}

Various strategies have been proposed to address the issue of data quality variability in mobile sensing applications. One common approach involves data preprocessing techniques, such as outlier removal~\cite{zhang2018platform} or data interpolation~\cite{yin2016mitigating}, which aim to alleviate the effects of poor-quality data. In addition to preprocessing, some studies explicitly consider varying data quality and incorporate it into the modeling process. For instance, Stisen \textit{et al.} train classifiers for datasets exhibiting diverse quality levels~\cite{stisen2015smart}. Khan \textit{et al.} employ domain adaptation techniques to adapt models to different contexts (e.g., user, device type, device instance)~\cite{khan2018scaling}. Chuprov \textit{et al.}~\cite{chuprov2022multi} design a Genetic Algorithm (GA)-based approach for sensor selection and fusion. Our research represents an initial effort towards creating a quality-aware mobile sensing system for earthquake detection, with a focus on analyzing accelerometer data quality on a large scale and assessing the effects of data quality variability. We acknowledge the emergence of novel methods in this area and intend to explore their potential application in future work.


\section{MyShake System and Dataset Overview}

\subsection{MyShake System}

MyShake is a global, smartphone-based seismic network that collects data from accelerometers to detect earthquakes and potentially provide earthquake early warning (EEW) using crowdsourced information~\cite{kong2016myshake, allen2020}. The MyShake app employs a trained Artificial Neural Networks (ANNs) model to identify earthquake-like movements on individual phones, only when the phone is stationary. If such movements are detected on a phone, a trigger (including location, time, and amplitude) is sent to a cloud server, where triggers from multiple devices are compiled and a network detection algorithm is used to confirm an earthquake~\cite{kong2020toward1}. In addition to each trigger, 3-axis accelerometer data is collected (as shown in Fig.~\ref{fig:examples}). Specifically, 5-minute segments (1 minute before and 4 minutes after the trigger) of 3-component acceleration data are recorded and uploaded to the cloud server when the phone has access to WiFi and power. The system is designed to sample accelerometer data at 25 Hz, resulting in approximately a 40-msec time interval between two samples. From all the uploaded waveforms, an earthquake waveform database is established~\cite{kong2019assessing}. Figure 1 illustrates the system architecture of MyShake.

\begin{figure}[tb]
	\centering
	\includegraphics[width=0.45\textwidth]{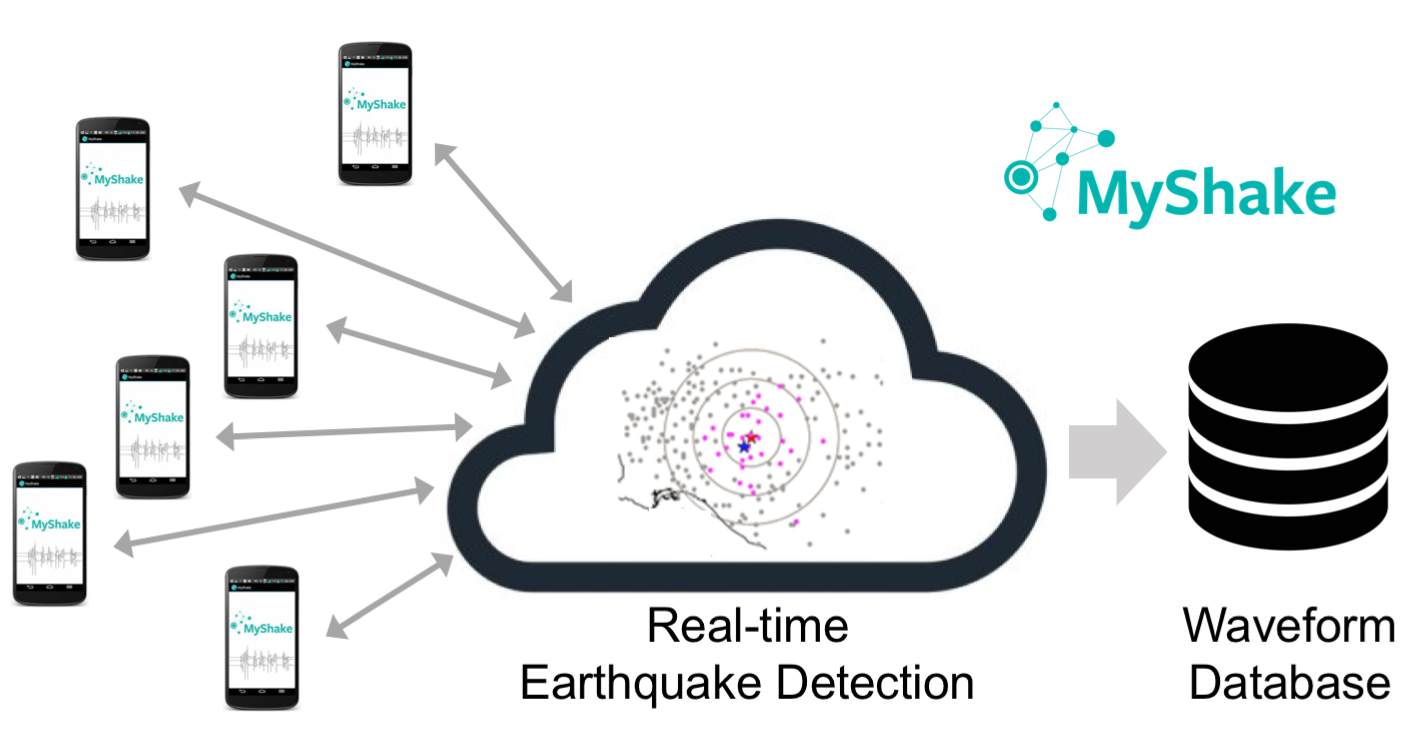}
	\caption{MyShake system architecture}
	\label{fig:architecture}
\end{figure}


\subsection{Dataset Description}

\begin{figure}[tb]
	\centering
	\includegraphics[width=0.48\textwidth]{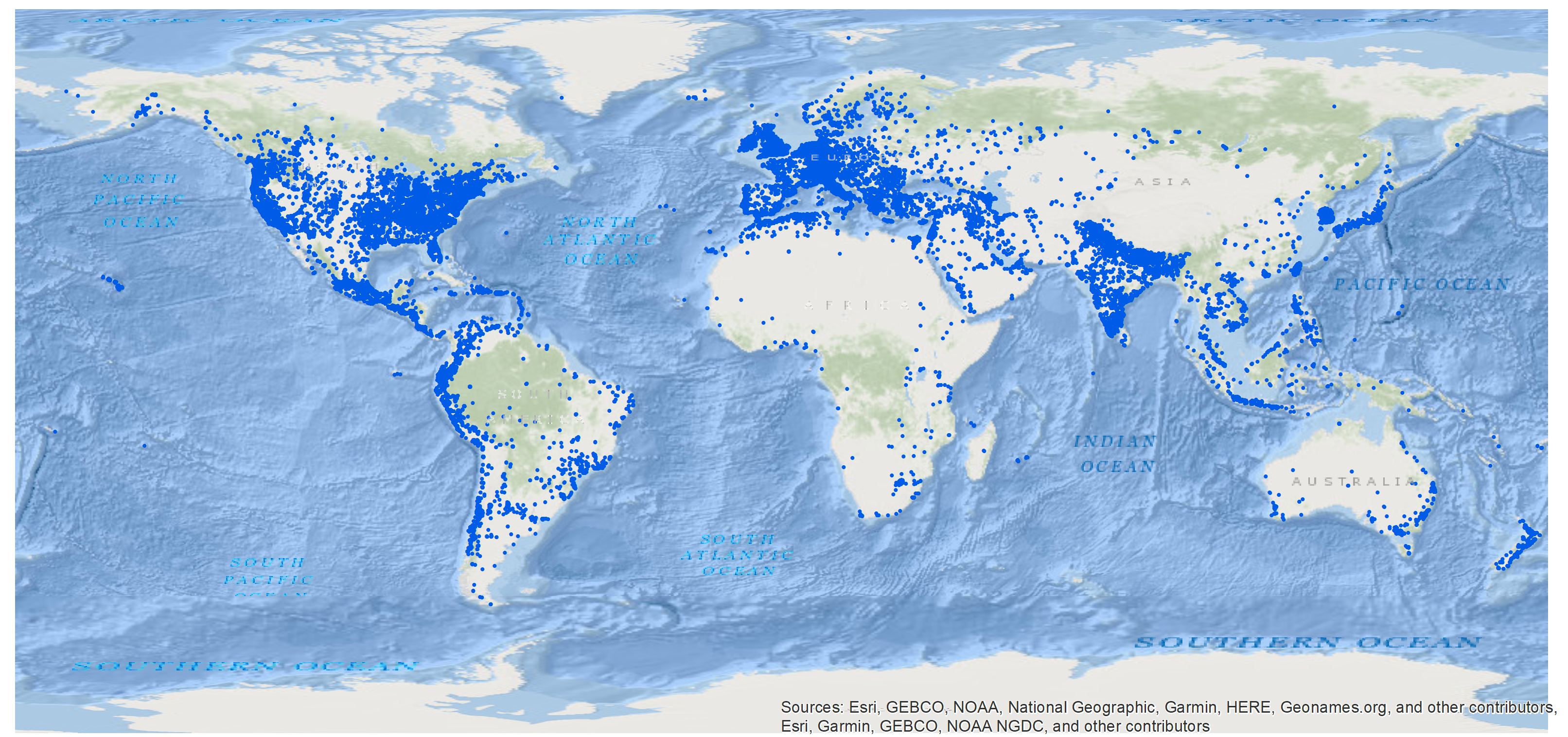}
	\caption{Spatial distribution of MyShake devices (with 1 km random spatial noise)}
	\label{fig:device}
\end{figure}

\begin{figure}
        \centering
		\begin{subfigure}[h]{0.43\textwidth} 
		\includegraphics[width=\textwidth]{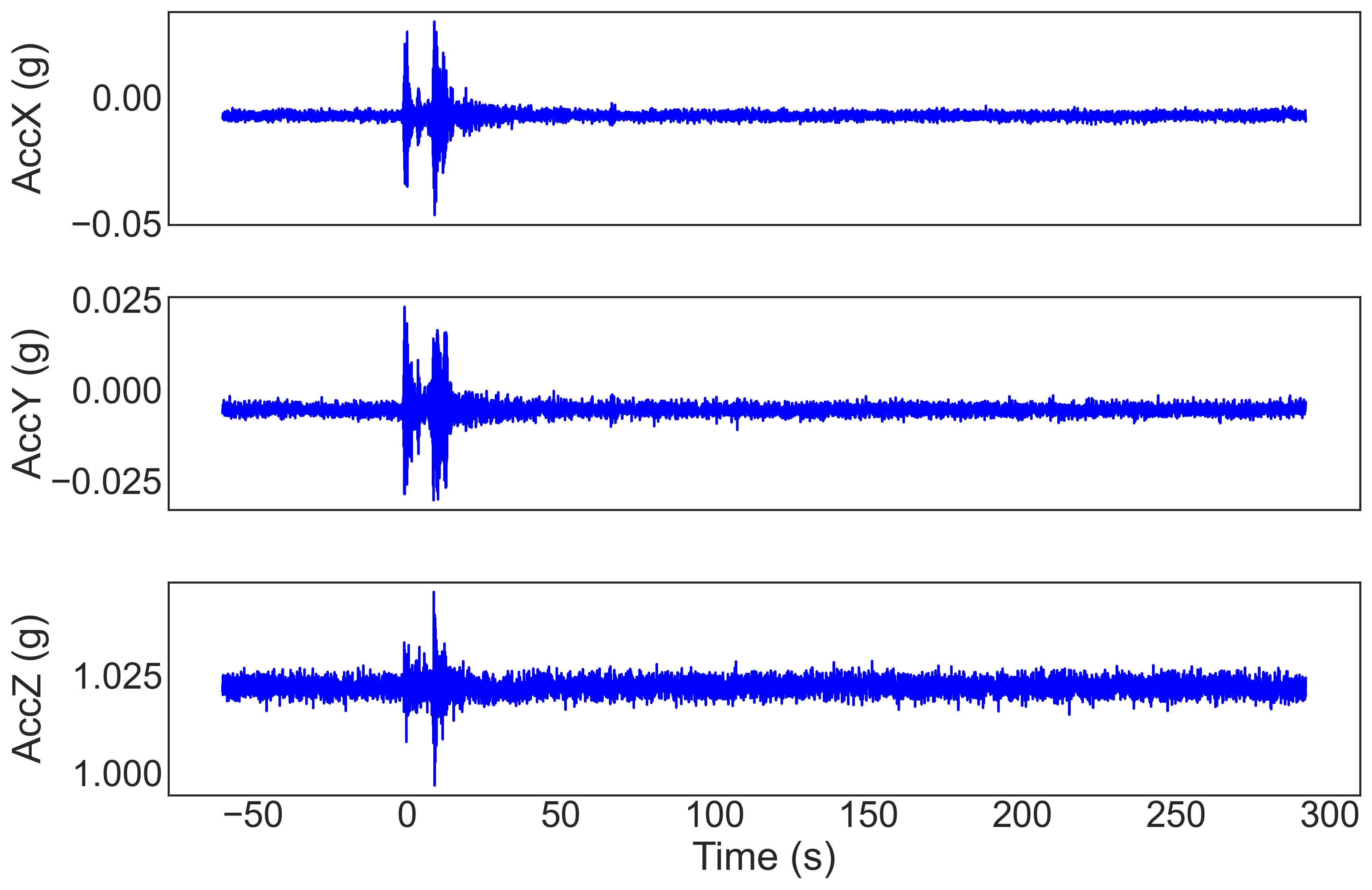}
		\caption{Good-quality waveforms}
		\label{fig:example_good}
		\end{subfigure}
		\begin{subfigure}[h]{0.43\textwidth} 
		\includegraphics[width=\textwidth]{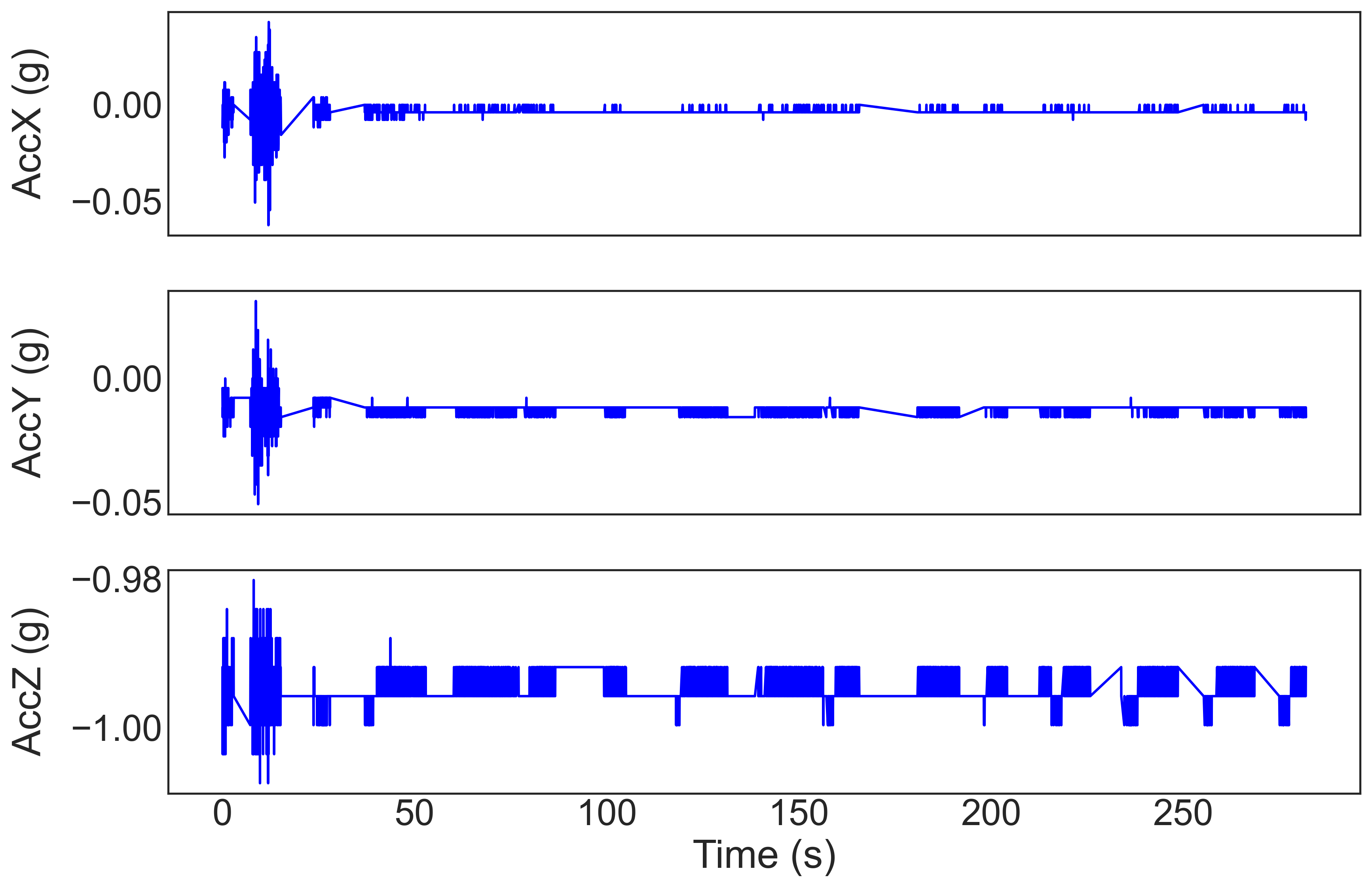}
		\caption{Waveforms with missing data}
		\label{fig:example_undersampling}
	  	\end{subfigure}
		\begin{subfigure}[h]{0.43\textwidth}  
		\includegraphics[width=\textwidth]{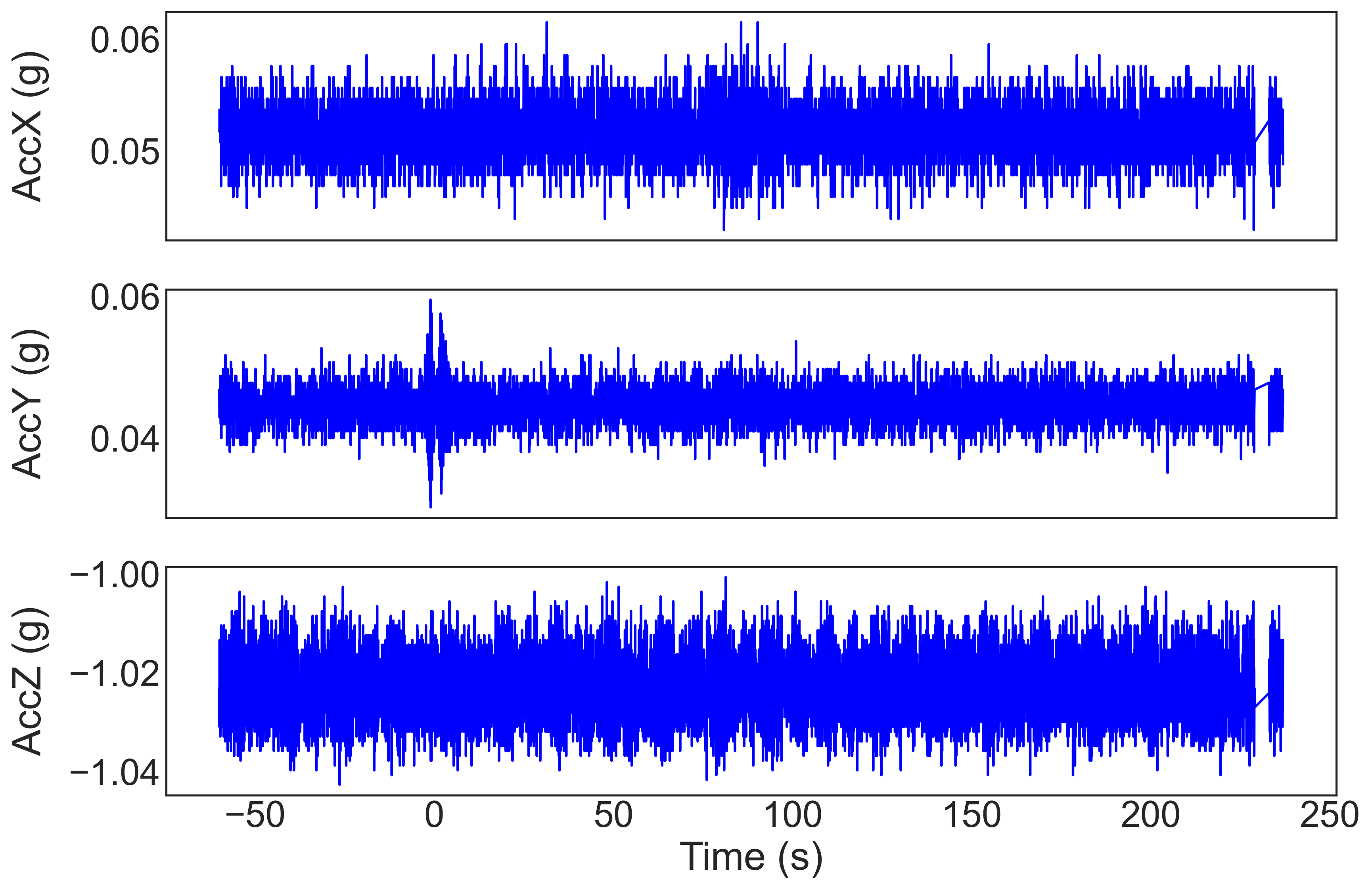}
		\caption{Waveforms with high noise levels}
		\label{fig:example_noise}
		\end{subfigure}
		\begin{subfigure}[h]{0.43\textwidth}  
		\includegraphics[width=\textwidth]{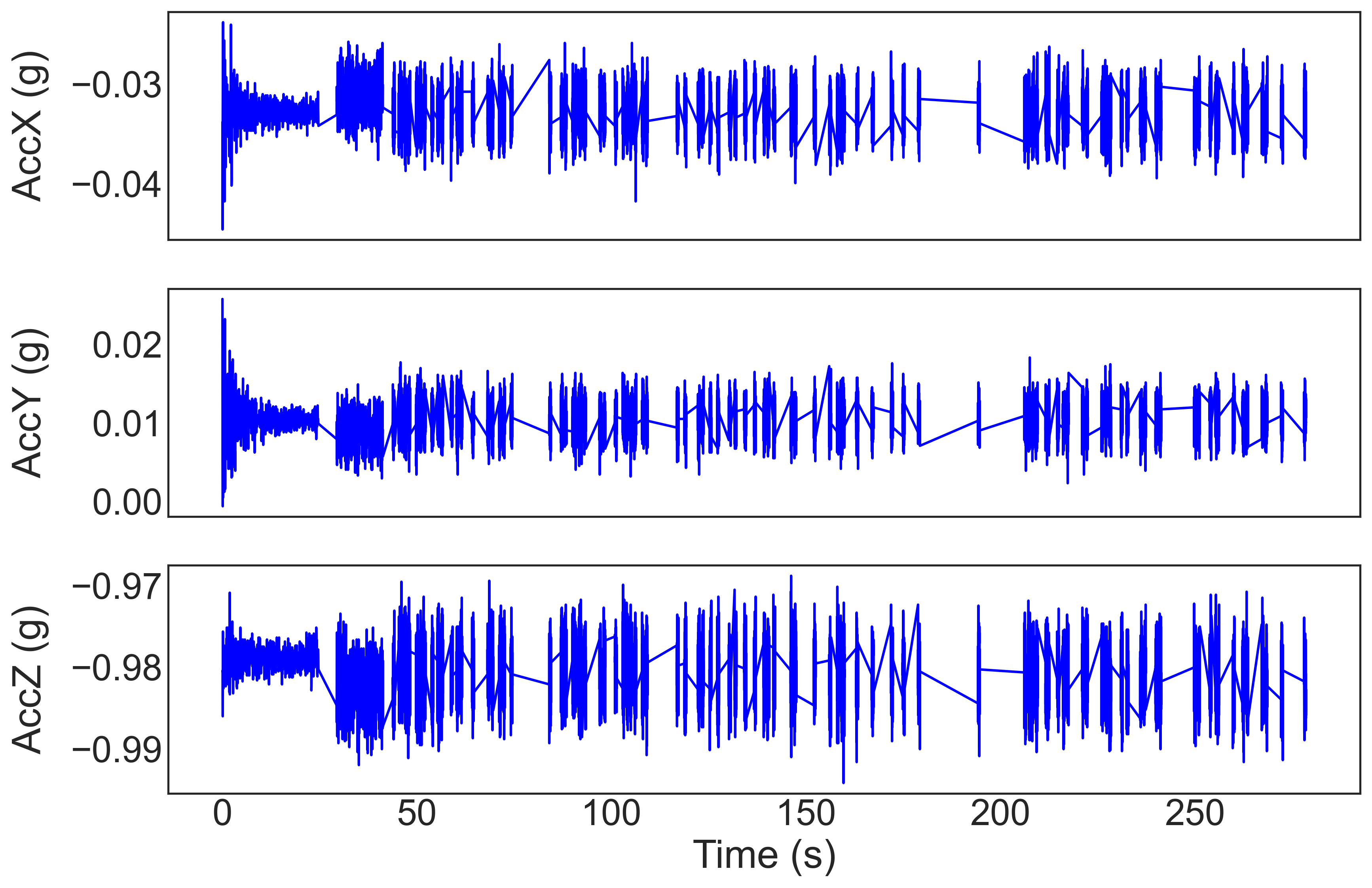}
		\caption{Waveforms with missing data and high noise levels}
		\label{fig:example_mix}
		\end{subfigure}
	\caption{Example waveforms with varying quality (Note: spikes near 0s represent earthquake-related triggers).} 
	\label{fig:examples}
\end{figure}

The MyShake earthquake waveform database provides the accelerometer dataset, comprising approximately 22 million anonymized waveforms collected from 81 thousand MyShake devices globally (all android devices), as depicted in Fig.~\ref{fig:device}. These waveforms were collected between 2016 and 2019. Information about the phone brands and accelerometer sensor vendors for these devices was also collected. MyShake obtains user consent to gather their phones' GPS locations and adds 1 km of random noise to the locations to safeguard user privacy. All 81 thousand devices include GPS location information.

\section{Accelerometer Data Quality}
\label{quality}

In this section, we firstly examine sample waveforms of varying quality. Secondly, we introduce the quality metrics employed to evaluate waveform data quality and present an analysis of the overall quality distribution.

\subsection{Example Waveforms}

Fig.~\ref{fig:examples} displays four sample waveforms with varying quality. The first example in Fig.~\ref{fig:example_good} exhibits good quality as it consistently collects samples within the 5-minute duration, and the 3-axis acceleration variations are very small before the earthquake-related trigger (near 0s). The second example in Fig.~\ref{fig:example_undersampling} presents noticeable missing data (i.e., gaps between time intervals) and a significantly smaller total number of samples compared to the example in Fig.~\ref{fig:example_good} (2773 vs. 8826). Fig.~\ref{fig:example_noise} showcases waveforms with high noise levels, as the acceleration values exhibit much larger variations than those in Fig.~\ref{fig:example_good}. In the y-axis, the noise-induced accelerations almost overshadow the earthquake-related accelerations. In Fig.~\ref{fig:example_noise}, the standard deviations of the noise level are substantially larger than those in Fig.\ref{fig:example_good} (0.0022 g vs. 0.0006 g). The example in Fig.~\ref{fig:example_mix} combines the issues observed in Fig.~\ref{fig:example_undersampling} and Fig.~\ref{fig:example_noise}, featuring both missing data and low resolution which we start to see the resolution levels.

\subsection{Quality Metrics}

Based on the analysis of example waveforms, we characterize waveform data quality using various metrics focused on sampling rate and noise level, which are highly relevant to earthquake detection performance. Sampling rate-based quality metrics have been extensively employed in other accelerometer-based applications as well~\cite{stisen2015smart,janidarmian2017comprehensive}. Specifically, we generate the following quality metrics:

\begin{itemize}
    \item \textbf{Sampling rate}: This includes the total number of samples (\textit{n\_sample}), the total number of pre-trigger samples (\textit{n\_noise}), and the standard deviation of time intervals between samples (\textit{std\_dt}).
    \item \textbf{Noise level}: This includes the standard deviation of noise data for the x, y, and z components (\textit{std\_x}, \textit{std\_y}, \textit{std\_z}). Note that for the vertical component, gravity is removed.
\end{itemize}

We calculate six quality metrics for each 5-minute segment of 3-axis acceleration data. The sampling rate affects the number of valid data points collected for earthquake detection. Based on the system's design, the expected values for \textit{n\_sample} and \textit{n\_noise} should be 7500 and 1500, respectively. \textit{std\_dt} serves as a measure of the sampling rate stability, with larger \textit{std\_dt} values indicating less stable time intervals throughout the recording period. The noise level can impact the accuracy of the recorded 3-axis acceleration data, which is crucial for estimating earthquake parameters. \textit{std\_x/y/z} provide a comparative measure of each device's noise level.





\subsection{Waveform Quality Distribution}

\begin{figure}[tb]
	\centering
	\includegraphics[width=0.5\textwidth]{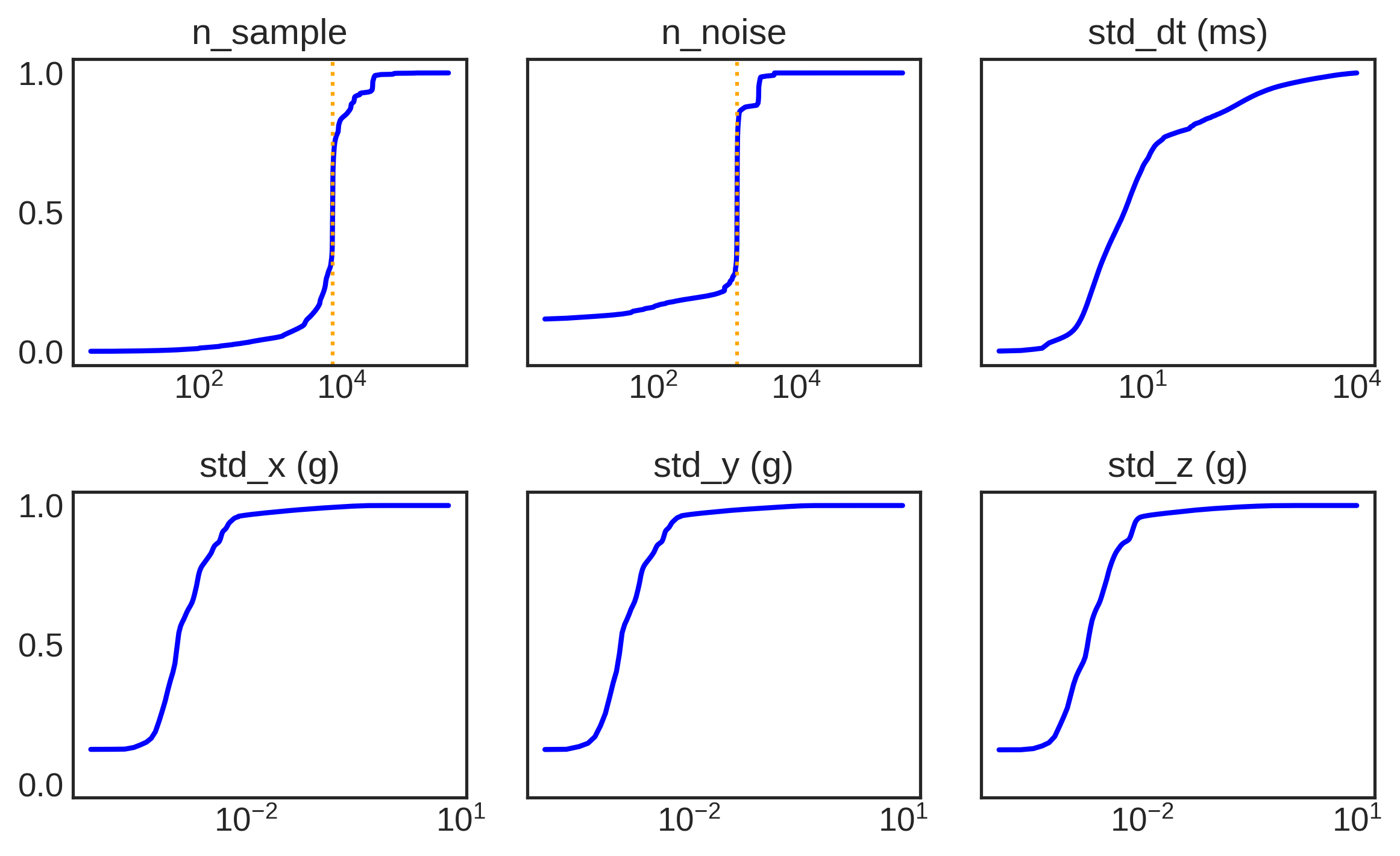}
	\caption{Cumulative distribution function (CDF) of six waveform data quality metrics, with dashed orange lines representing the expected values of \textit{n\_sample} and \textit{n\_noise}. This figure highlights significant variations observed in waveform quality distributions. }
	\label{fig:quality_metrics}
\end{figure}

Fig.~\ref{fig:quality_metrics} presents the cumulative distributions of six waveform quality metrics. Although the MyShake system is designed to sample data at 25 Hz, numerous waveforms display variations in terms of sampling rate. Regarding \textit{n\_sample}, approximately half of the waveforms contain around 7500 samples. For the remaining waveforms, there is a higher percentage of undersampling instances (31.1\%) compared to oversampling instances (24.7\%). The patterns of \textit{n\_noise} are roughly similar to those of \textit{n\_sample}. \textit{std\_dt} ranges from 0 to 10,000 msec, suggesting the presence of varying gap sizes between samples. As for noise level, \textit{std\_x}, \textit{std\_y}, and \textit{std\_z} exhibit comparable distributions. Most variations fall between 0.0005 and 0.01 g, but a very small portion of waveforms still demonstrate extremely large noise variations. 

\section{Factors related to Accelerometer Data Quality}
\label{factor}

In this section, we explore potential factors influencing accelerometer-based sensing quality, such as phone and accelerometer manufacturer, accelerometer model, phone specifications, geolocation, and trigger time. It is important to note that, in addition to these observable factors, numerous unobserved factors like user behavior and local environment can also affect accelerometer data quality. Our study, however, primarily focuses on the observed factors.

\subsection{Smartphone Hardware}

For each MyShake device, both smartphone and accelerometer sensor information was collected. Smartphone information consists of its manufacturer and specific model, for example, \textit{``samsung, galaxy a7''}. Accelerometer sensor information consists of its manufacturer and specific model, for example, \textit{``st, lsm330''} (``st'' is short for STMicroelectronics, which is a well-known semiconductor manufacturer). \bigskip

The raw information collected can be noisy, inconsistent, or even incorrect. Therefore, we perform preprocessing on the smartphone and accelerometer sensor information. After preprocessing, there are 513 phone manufacturers, 4036 phone models, 17 accelerometer manufacturers, and 419 accelerometer models. Among the 513 phone manufacturers, only 26 manufacturers have at least 100 MyShake devices, and they account for 96.3\% of all devices (81 thousand in total). This suggests that most MyShake devices are associated with major players in the smartphone market. Consequently, we rank phone manufacturers and accelerometer manufacturers by the number of devices linked to them and examine the most representative ones. \bigskip

\begin{figure}[tb]
	\centering
	\includegraphics[width=0.5\textwidth]{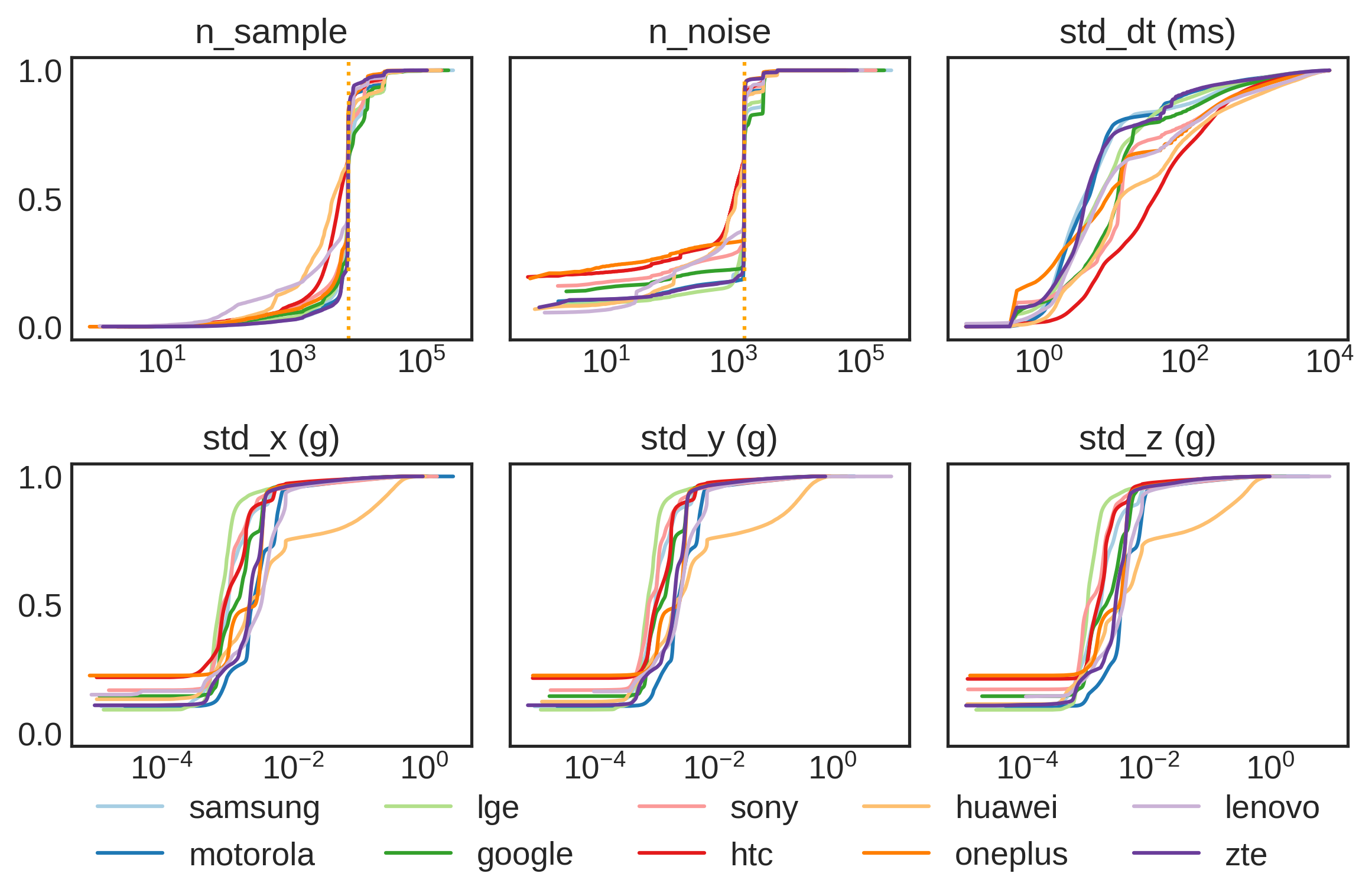}
	\caption{Cumulative distribution function (CDF) of waveform quality metrics for the top 10 smartphone manufacturers. This figure highlights that there are differences in sampling rates and noise levels across various phone manufacturers, with significant deviations observed in \textit{htc} and \textit{huawei} phones. } 
	\label{fig:quality_phone_manufacturers}
\end{figure}

Fig.~\ref{fig:quality_phone_manufacturers} displays the cumulative distributions of quality metrics for various phone manufacturers, revealing differences in both sampling rate and noise levels. Waveforms from certain phone manufacturers exhibit significant deviations compared to others. For example, the waveforms of \textit{htc} phones demonstrate a deviation in \textit{std\_dt}. The median \textit{std\_dt} for \textit{htc} phones is 39 msec, considerably larger than that of other manufacturers, such as \textit{samsung} at 4 msec. A larger \textit{std\_dt} implies greater variations in time intervals (i.e., an unstable sampling rate). Another notable deviation is found in \textit{huawei}, which displays different distributions in \textit{std\_x}, \textit{std\_y}, and \textit{std\_z}, suggesting that waveforms from \textit{huawei} phones are more likely to exhibit high noise levels. These significant deviations point to potential quality issues in the waveforms collected by those phone manufacturers. \bigskip

\begin{figure}[tb]
	\centering
	\includegraphics[width=0.5\textwidth]{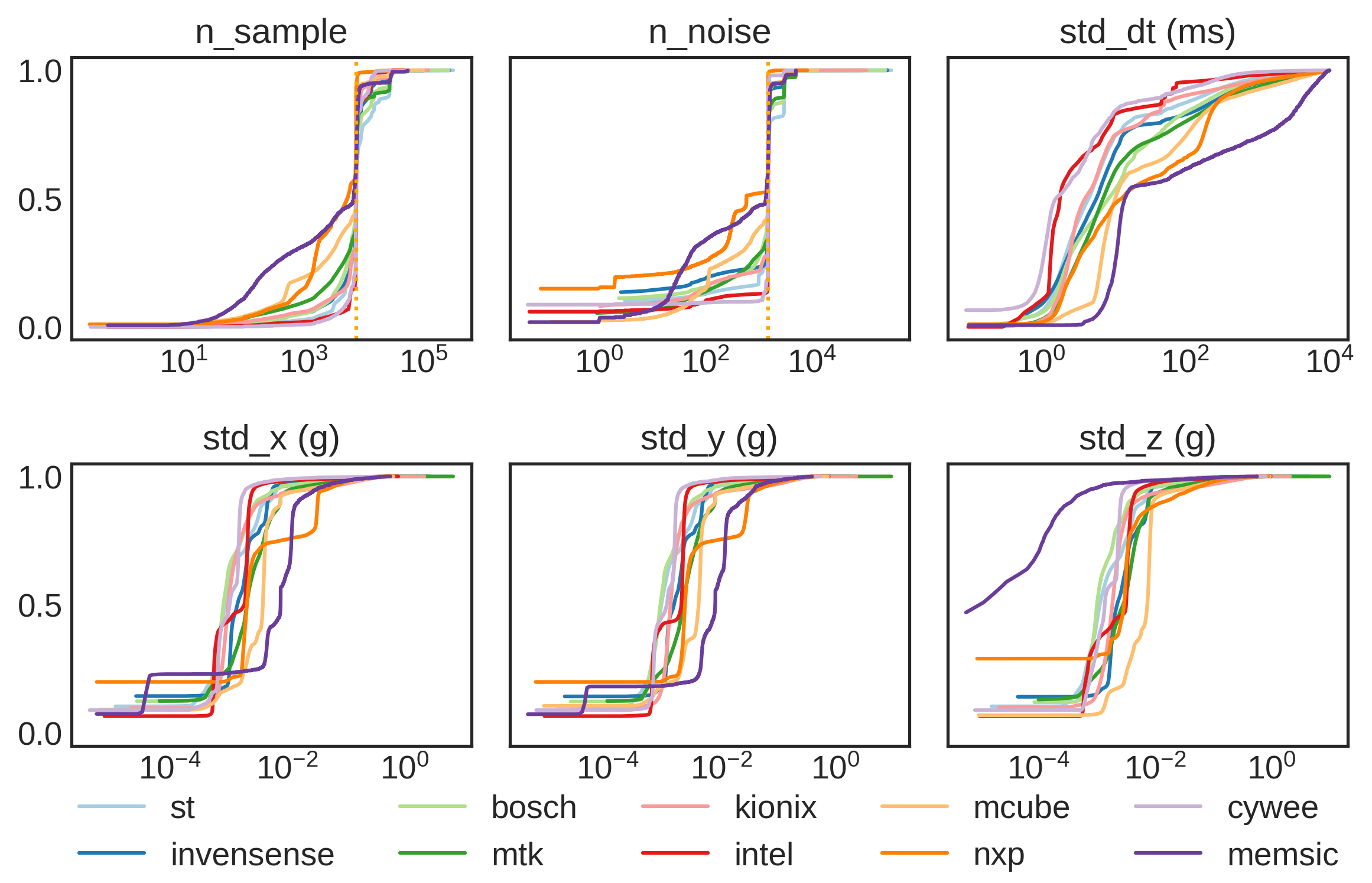}
	\caption{Cumulative distribution function (CDF) of waveform quality metrics for the top 10 accelerometer manufacturers. This figure highlights that accelerometer manufacturers show greater variations in sampling rate and noise levels than phone manufacturers. } 
	\label{fig:quality_acc_manufacturers}
\end{figure}
\begin{figure}[tb]
	\centering
	\includegraphics[width=0.48\textwidth]{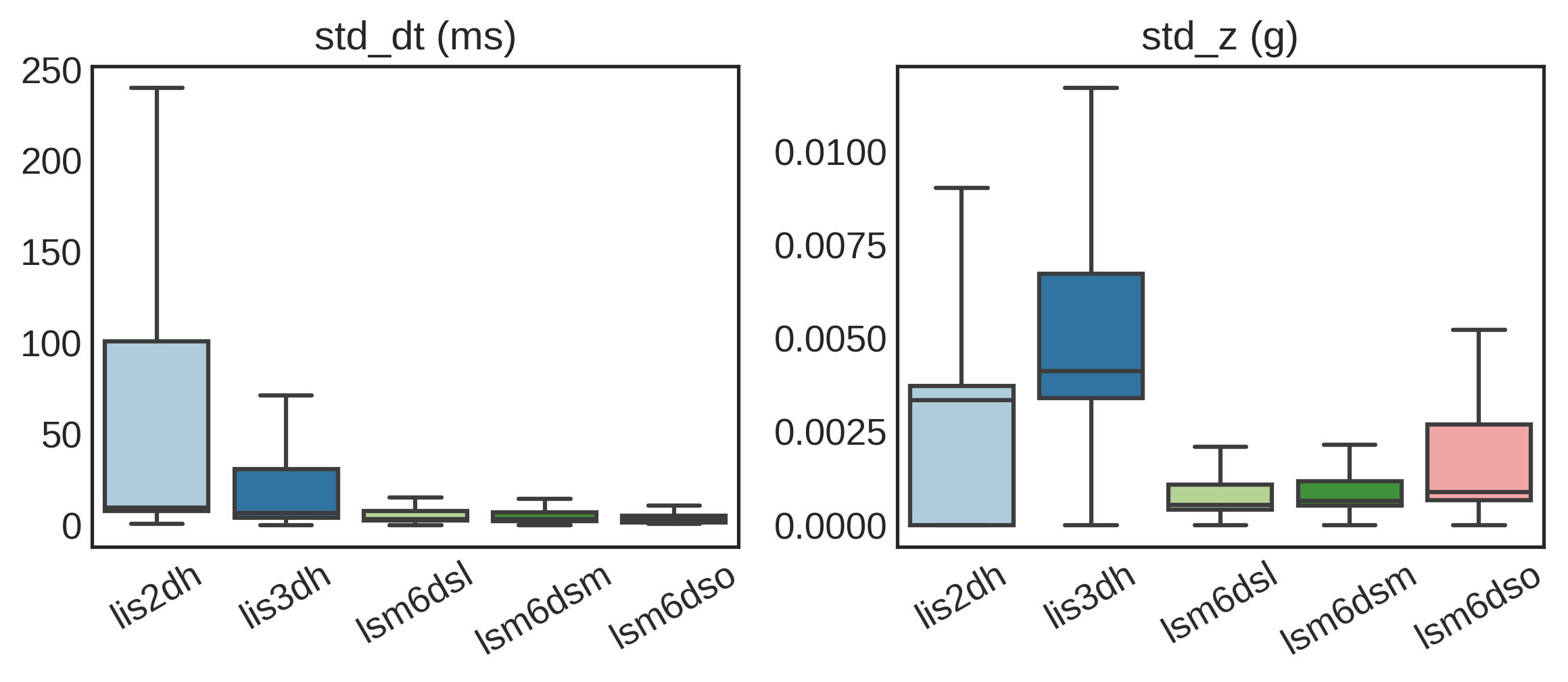}
	\caption{Boxplots of waveform quality metrics across various STMicroelectronics accelerometer models (ordered by release time). This figure highlights that even among different models from the same accelerometer manufacturer, such as STMicroelectronics, variations in sampling intervals and noise levels exist, with newer models generally exhibiting smaller variations in time intervals and noise levels.} 
	\label{fig:quality_acc_version}
\end{figure}

Fig.~\ref{fig:quality_acc_manufacturers} presents the cumulative distributions of quality metrics for various accelerometer manufacturers, revealing greater variations in both sampling rate and noise levels compared to phone manufacturers. Regarding sampling rate, variations among accelerometer manufacturers are primarily observed in the undersampling aspect. Manufacturers such as \textit{memsic}, \textit{mcube}, and \textit{nxp} exhibit a higher proportion of undersampling cases compared to others. Additionally, waveforms from \textit{memsic} show significant deviations in the distributions of \textit{std\_x}, \textit{std\_y}, and \textit{std\_z}. Fig.~\ref{fig:quality_acc_version} illustrates that even waveforms from the same accelerometer manufacturer but different models display variations in sampling interval and noise level. In this example, all five models are from STMicroelectronics and are ordered by release time. Newer generation models like \textit{lsm6dsl}, \textit{lsm6dsm}, and \textit{lsm6dso} exhibit smaller variations in time intervals and noise levels compared to older models such as \textit{lis2dh} and \textit{lis3dh}. Interestingly, newer models within the same series tend to have slightly larger variations in noise levels (e.g., \textit{lis3dh} vs. \textit{lis2dh}).


\subsection{Smartphone Specifications}


Due to the extensive number of phone models, examining the main differences between all of them can be challenging. To address this issue, we gather specifications for each phone model using their brand and model information from a public website called GSMArena~\footnote{www.gsmarena.com}, which provides detailed smartphone specifications. The queried specifications include release date, random access memory (RAM) size, and battery size. Out of the 4036 phone models, we successfully collected phone specifications for 1663 models.

\begin{figure}[tb]
	\centering
	\includegraphics[width=0.5\textwidth]{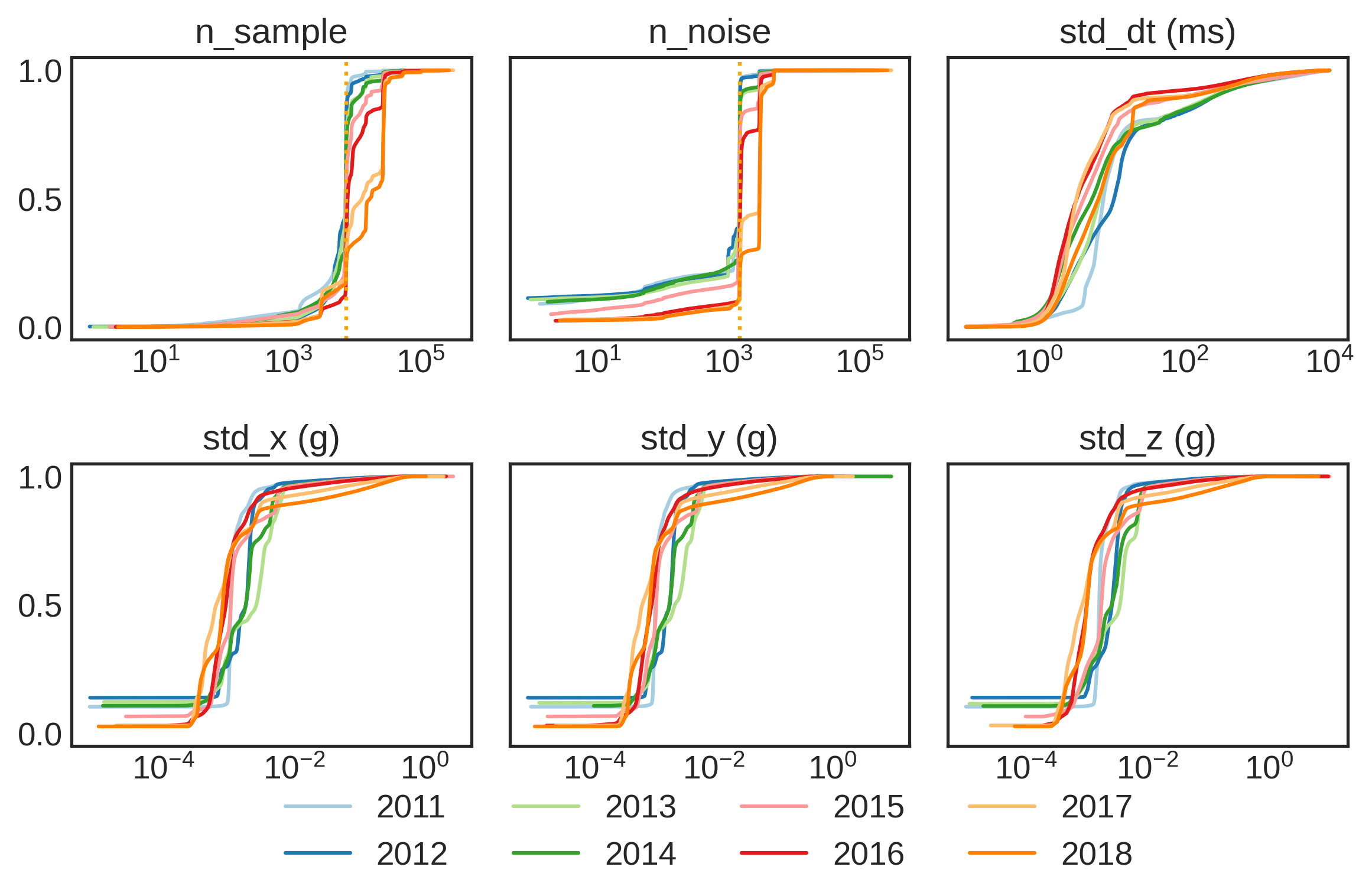}
	\caption{Cumulative distribution function (CDF) of waveform quality metrics categorized by the release year of phones. This figure highlights that newer phones are more likely to oversample.} 
	\label{fig:quality_model_year}
\end{figure}
\begin{figure}[tb]
	\centering
	\includegraphics[width=0.5\textwidth]{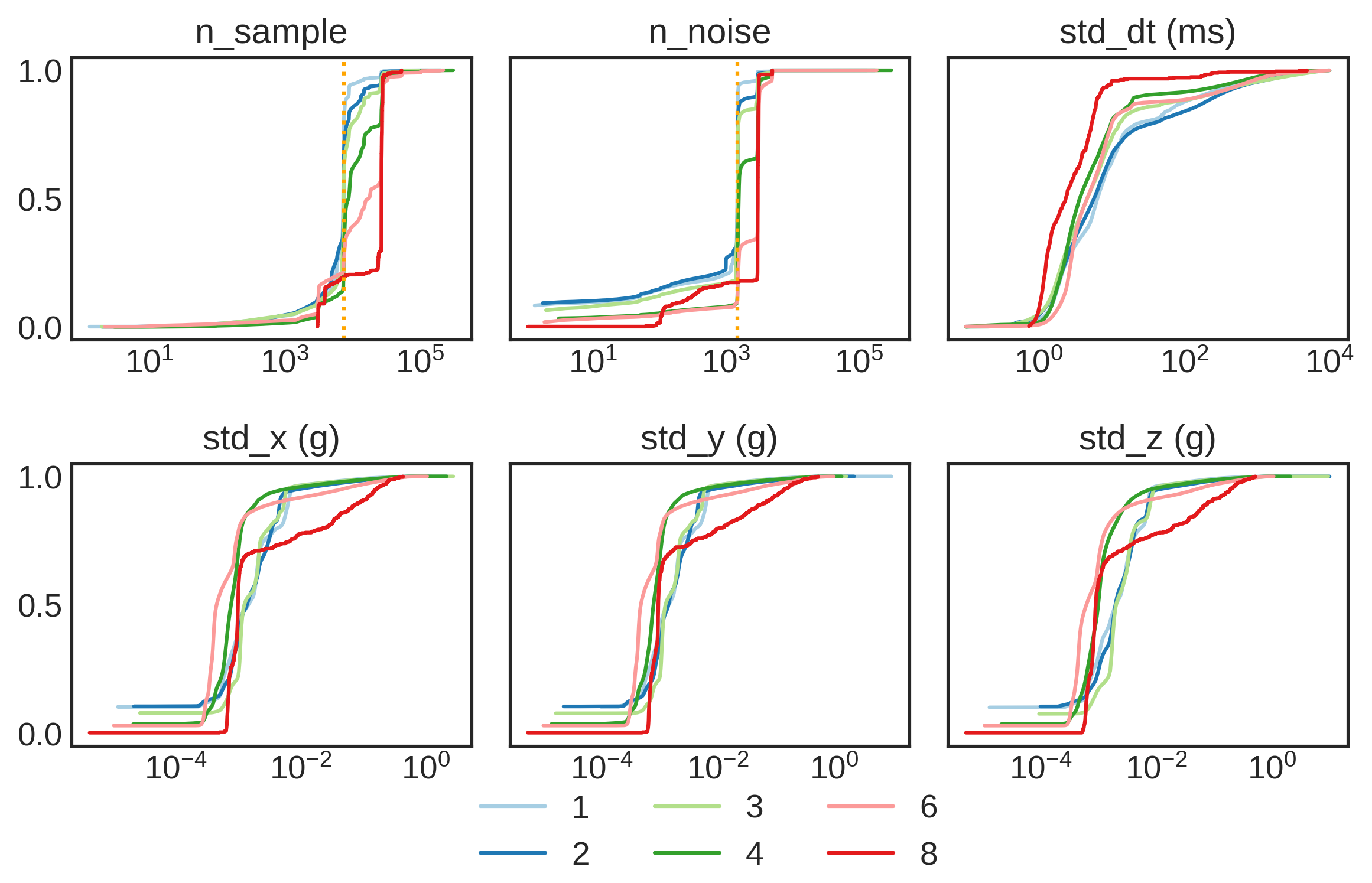}
	\caption{Cumulative distribution function (CDF) of waveform quality metrics categorized by phone RAM size (unit: GB). This figure highlights that phones with larger RAM are more likely to oversample. } 
	\label{fig:quality_model_RAM}
\end{figure}

Fig.~\ref{fig:quality_model_year} displays the cumulative distributions of quality metrics for phones released in different years, revealing variations in \textit{n\_sample} and \textit{n\_noise}. A noticeable pattern emerges, showing that newer phones are more likely to oversample. Fig.~\ref{fig:quality_model_RAM} presents the cumulative distributions of quality metrics for phones with different RAM sizes, demonstrating similar patterns to release year: phones with larger RAM are more likely to oversample. Notably, the 8 GB RAM size group deviates significantly in \textit{std\_x}, \textit{std\_y}, and \textit{std\_z}, suggesting that waveforms in this group tend to have larger variations in noise levels. Additionally, phones with different battery sizes exhibit smaller variations in sampling rate compared to release year and RAM size, indicating a lesser impact of battery size on sampling rate. Regarding noise level, battery sizes below 5000 mAh display similar distributions, while those above 5000 mAh are less likely to exhibit high noise levels.

\subsection{Geolocation}

The GPS location of MyShake devices in this study represents a single snapshot in time (with 1 km random noise added). We associate each GPS location (latitude, longitude) with geographic layers, such as country boundaries and roads~\footnote{https://www.naturalearthdata.com/downloads/10m-cultural-vectors/roads/}, to obtain the geographic context (e.g., country, distance to highway) of the MyShake devices. \bigskip

\begin{figure}[tb]
	\centering
	\includegraphics[width=0.4\textwidth]{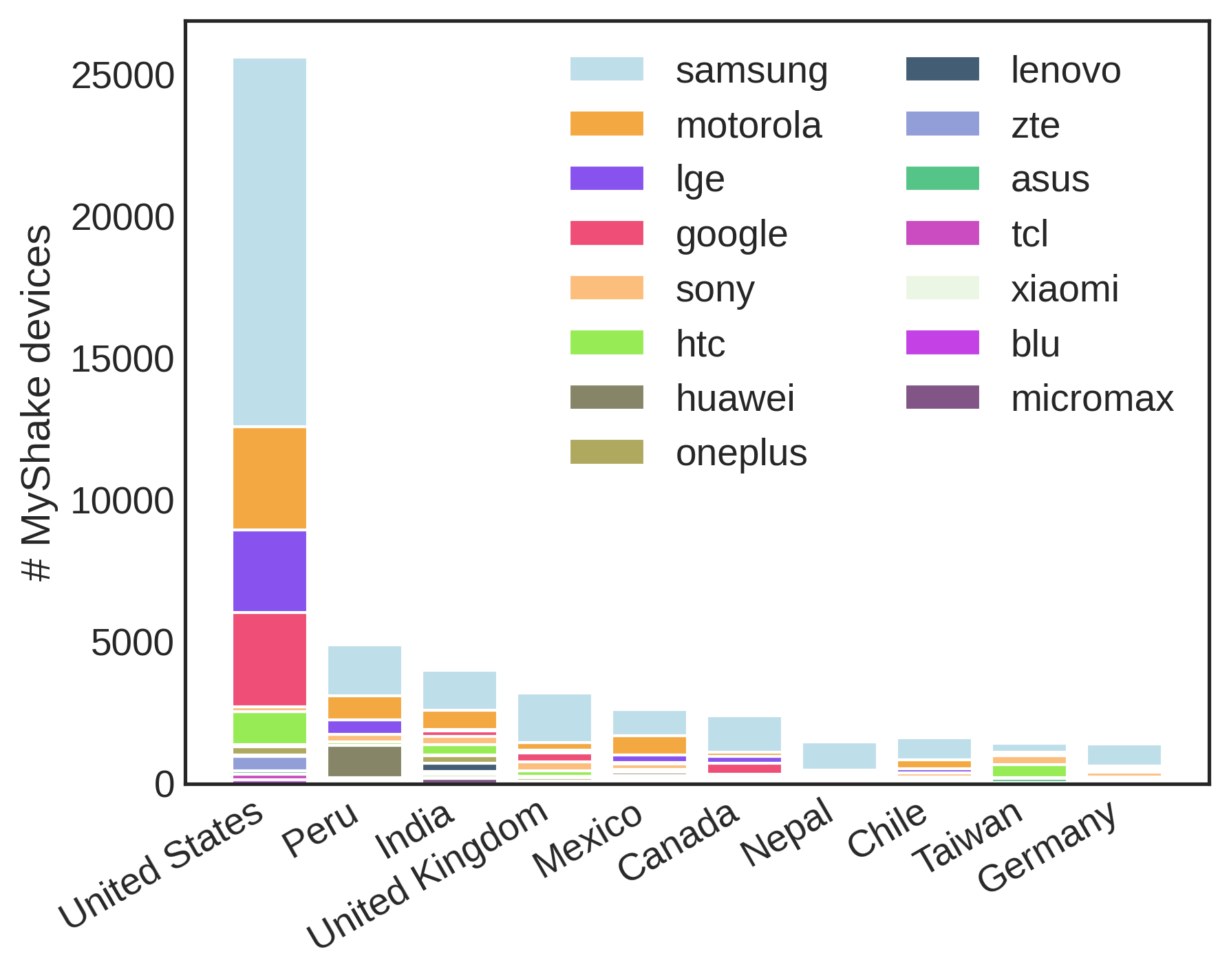}
	\caption{Phone manufacturer distribution in the top 10 countries with MyShake devices. This figure highlights the varying compositions of phone manufacturers across countries. } 
	\label{fig:top_country_brand}
\end{figure}

\begin{figure}[tb]
	\centering
	\includegraphics[width=0.4\textwidth]{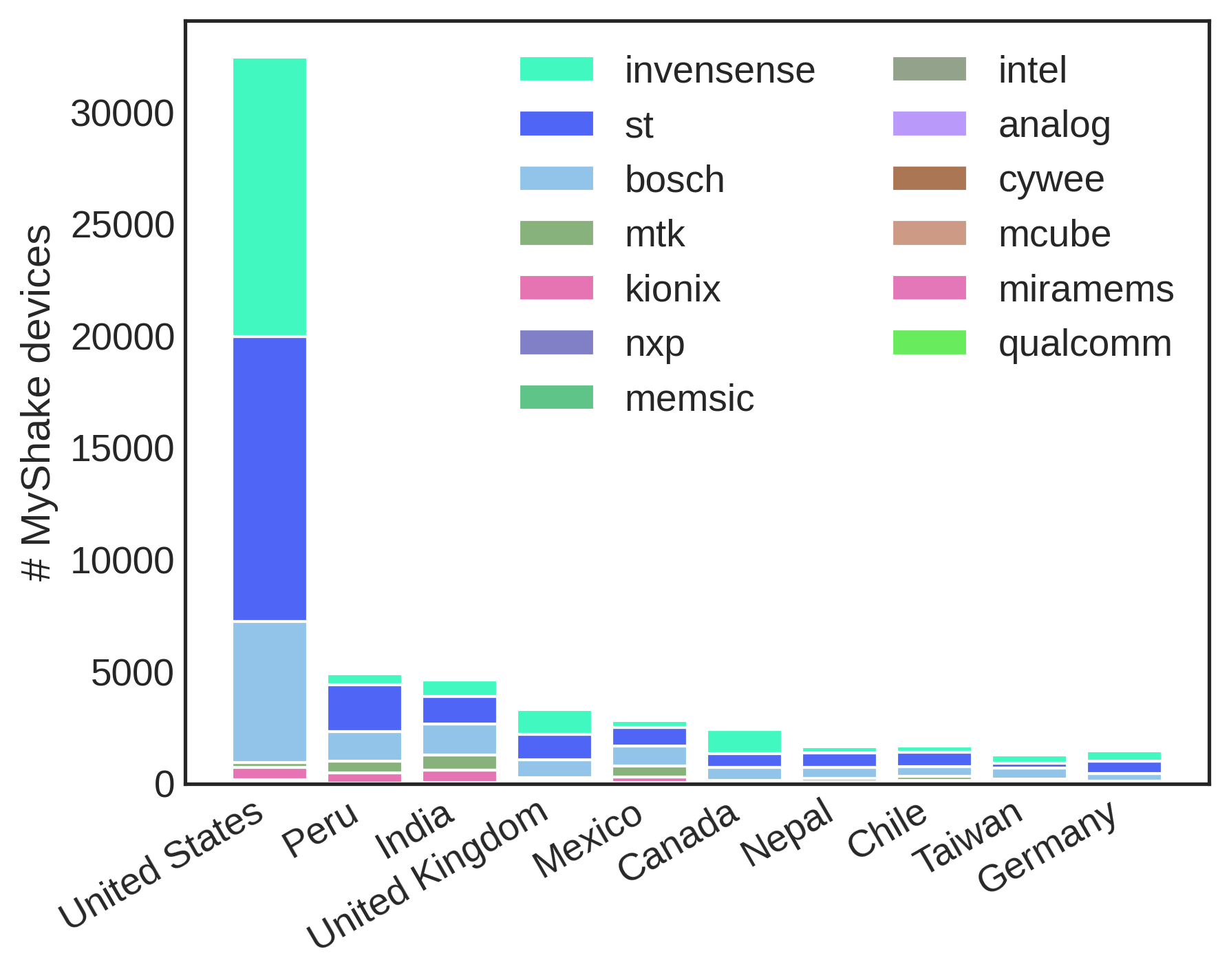}
	\caption{Accelerometer manufacturer distribution in the top 10 countries with MyShake devices. This figure highlights that the top 10 countries with MyShake devices have most of their accelerometer sensors from three manufacturers (\textit{invensense}, \textit{st}, and \textit{bosch}), with varying ratios. } 
	\label{fig:top_country_acc}
\end{figure}

\begin{figure}[tb]
	\centering
	\includegraphics[width=0.49\textwidth]{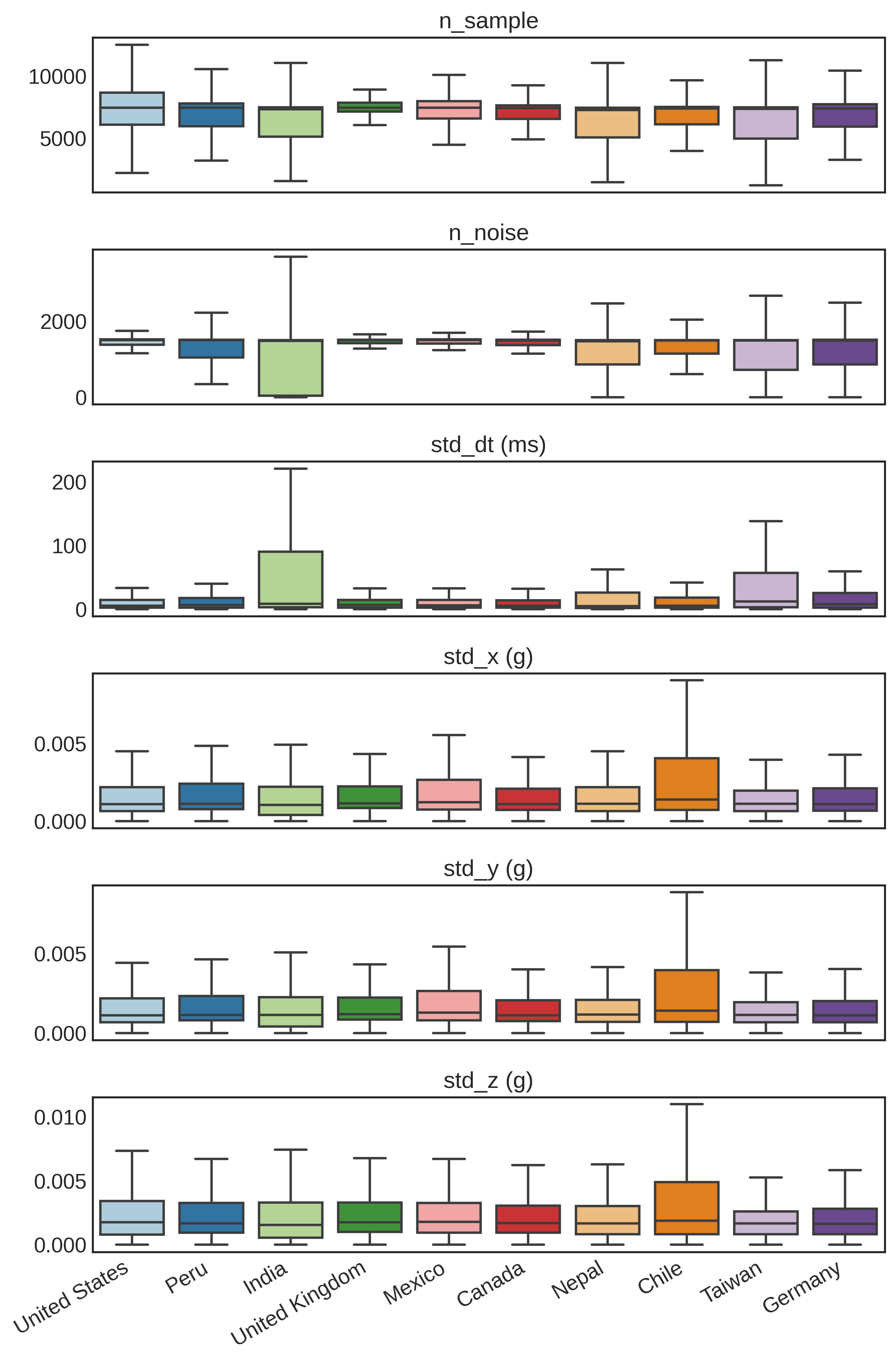}
	\caption{Boxplots of waveform quality metrics for the top 10 countries with MyShake devices. This figure highlights that waveforms from \textit{India}, \textit{Nepal}, and \textit{Taiwan} show a higher likelihood of undersampling, while most countries exhibit similar noise level distributions, except for \textit{Chile}, which has larger variations in noise levels.} 
	\label{fig:quality_country}
\end{figure}

Using the country information, we calculate the total number of MyShake devices per country and select the top 10 countries. We then examine their hardware characteristics in terms of phone and accelerometer manufacturer composition. As illustrated in Fig.~\ref{fig:top_country_brand}, a significant percentage of phones from the top 10 countries are \textit{samsung} devices, with varying compositions of phone manufacturers across countries. The United States and India have the largest number of phone manufacturers (105 and 89). Fig.~\ref{fig:top_country_acc} displays the accelerometer manufacturer compositions, with most devices in the top 10 countries featuring sensors from three manufacturers: \textit{invensense}, \textit{st}, and \textit{bosch}, albeit with different ratios. There are fewer variations in accelerometer manufacturer compositions among the top 10 countries compared to phone manufacturer compositions. As different countries have varying compositions of phone and accelerometer manufacturers, they also display differences in waveform quality. Fig.~\ref{fig:quality_country} shows that waveforms from \textit{India}, \textit{Nepal}, and \textit{Taiwan} are more likely to be undersampled and have larger time intervals compared to other countries. Meanwhile, most countries have similar noise level distributions, with \textit{Chile} standing out as having larger variations in noise levels. \bigskip

 

\begin{figure}[tb]
	\centering
	\includegraphics[width=0.48\textwidth]{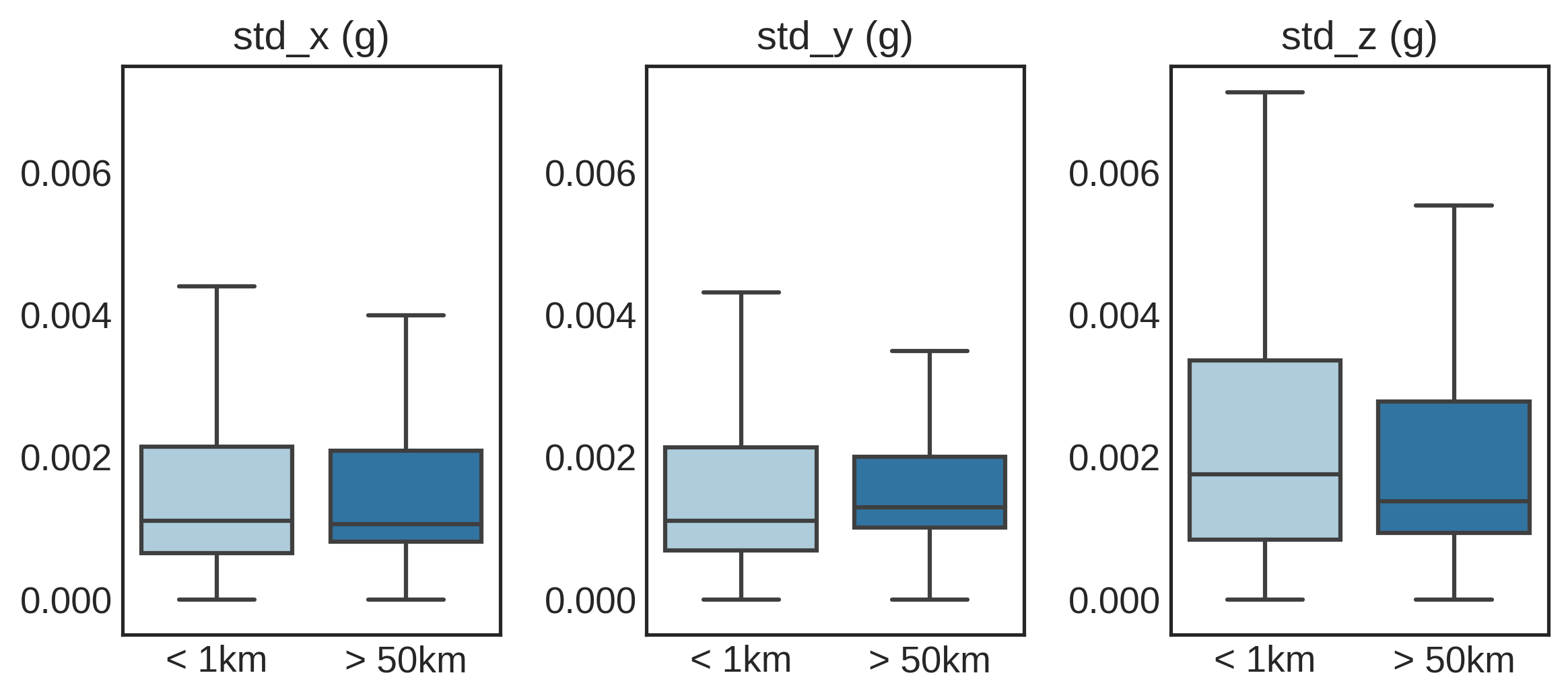}
	\caption{Boxplots of waveform quality metrics of devices very close to (within 1km) and farther away from (beyond 50km) highways. This figure highlights the difference in noise levels between devices near highways and those farther away, with the latter group exhibiting less noise variations. } 
	\label{fig:quality_highway}
\end{figure}

To assess the influence of local environments on ambient noise levels, we explore an additional geographic factor, specifically the proximity to highways. We focus on devices within the United States and calculate the distance of each device to the nearest primary and secondary highways. We then compare devices situated very close to highways (within 1km) to those located at a considerable distance (beyond 50km) from highways. As illustrated in Fig.~\ref{fig:quality_highway}, there is a notable disparity in noise levels between the two groups, with devices farther away from highways exhibiting relatively less variation in noise levels. This is particularly evident for the z-axis, where devices situated farther from highways demonstrate a lower median \textit{std\_z} compared to those in close proximity to highways. 




\subsection{Trigger Time}

\begin{figure}[tb]
	\centering
	\includegraphics[width=0.45\textwidth]{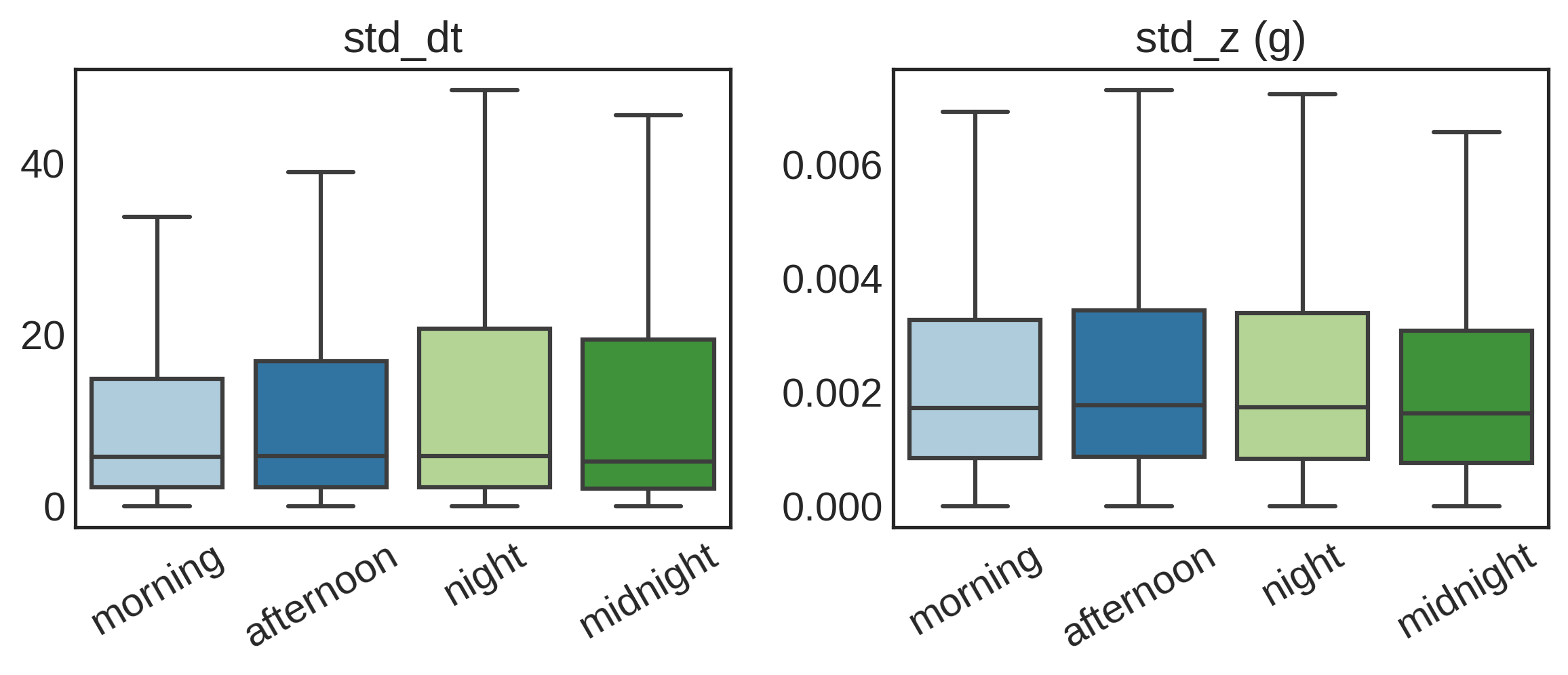}
	\caption{Cumulative distribution function (CDF) of waveform quality metrics categorized by trigger time period. This figure highlights that waveforms collected during afternoon and evening hours exhibiting larger time intervals and slightly higher noise levels.} 
	\label{fig:quality_period}
\end{figure}

The quality of waveform data may also be influenced by the time of collection, which is closely associated with users' daily routines and phone usage patterns. We determine the local hour for each waveform using its corresponding trigger timestamp. The variations in quality metrics across different hours are relatively minor, particularly when compared to hardware-related factors. Nonetheless, certain distinctions related to users' behavior can still be observed. In Fig.~\ref{fig:quality_period}, we evaluate waveform data quality across different time periods. Due to increased phone usage, waveforms gathered during the afternoon (12 - 6 pm) and evening (7 - 11 pm) exhibit larger time intervals and slightly higher noise levels compared to those collected during the morning (6 - 11 am).


\section{Factor Importance}

In this section, we predict waveform data quality by taking into account the potential impact factors identified earlier, with the aim of evaluating factor importance. Considering the availability of impact factors, we compile a dataset comprising waveforms that possess a complete set of features. This dataset includes approximately 10 million waveforms from 31 thousand MyShake devices, encompassing 845 phone models across 22 manufacturers, and 78 accelerometer models from 11 different manufacturers. Focusing on waveforms with quality issues, we develop classification models to identify these poor-quality waveforms. \bigskip

As an initial step, we establish rules for quality control, which serve to to distinguish between good- and poor-quality waveforms. Based on the distributions of quality metrics displayed in Fig.~\ref{fig:quality_metrics}, we calculate percentiles for each metric and combine them at various levels. Specifically, for a given quality level k\%, we compute the lower k\% values of \textit{n\_sample} and \textit{n\_noise}, and the higher k\% values of \textit{std\_dt}, \textit{std\_x}, \textit{std\_y}, and \textit{std\_z}. These values, which serve as thresholds, are then combined into a condition designed to identify waveforms exhibiting significant undersampling or high noise level issues. We test k values of 10, 15, 20, and 25, where higher k values will classify more waveforms as \textit{poor-quality}. \bigskip

For quality prediction, we divide the entire dataset into training (70\%) and test (30\%) sets. We employ a Random Forest binary classifier and utilize weight balancing to address the class imbalance issue (i.e., more good-quality than poor-quality waveforms). For categorical features, such as phone and accelerometer manufacturers, we convert them into numerical values using one-hot encoding. To assess classification performance, we rely on metrics such as precision, recall, and F1-score. Specifically, a \textit{true positive} (TP) occurs when the predicted poor-quality waveform is from the \textit{poor-quality} group; otherwise, it is a \textit{false positive} (FP). When a waveform from the \textit{poor-quality} group is not predicted as a poor-quality waveform by the model, it is considered a \textit{false negative} (FN). Using these definitions, we compute the evaluation metrics accordingly: 


\begin{equation}
    \small
    precision = \frac{|TP|}{|TP| + |FP|}
\end{equation}

\begin{equation}
    \small
    recall = \frac{|TP|}{|TP| + |FN|}
\end{equation}

\begin{equation}
    \small
    F1 = \frac{2}{precision^{-1} + recall^{-1}}
\end{equation}


Table~\ref{tb:prediction_level} presents the classification performance at varying quality control levels (k\%). As the quality control level increases, a larger number of waveforms are categorized into the \textit{poor-quality} group, making it easier for the model to detect them. The 25\% quality control level achieves the best prediction performance with an F1-score of 0.76. The 10\% level focuses on more extreme cases of undersampling and/or high noise levels, and the impact factors can still provide reasonable predictions for such cases. It is important to note that we do not expect a very high F1-score, as our prediction only includes observable features, while numerous unobservable features could also affect waveform quality.

\begin{table}[h]
    \small
	\centering
	\caption{Comparison of classification performances with different quality control (QC) levels. }
	\label{tb:prediction_level}
		\begin{tabular}{cccc}
			\toprule[1.2pt]
			\bf QC Level & \bf Precision & \bf Recall & \bf F1-score   \\ \midrule[0.5pt]
			\bf 10\% & 0.58 & 0.61 & 0.60 \\ 
			\bf 15\% & 0.65 & 0.61 & 0.63 \\ 
			\bf 20\% & 0.77 & 0.64 & 0.70 \\ 
			\bf 25\% & 0.84 & 0.70 & 0.76  \\ 
	    \midrule
		\end{tabular}
\end{table}



We further compare prediction performance using various feature sets, applying the 10\% quality control level to target extreme cases. Based on the prediction performances shown in Table~\ref{tb:prediction_feature}, the most important feature sets are accelerometer model and phone specifications, which include key information about the sensor and phone resources. The location feature set ranks as the third most important. Accelerometer and phone manufacturer features are less effective in predicting poor-quality waveforms, with the time feature being the least important. 

\begin{table}[h]
    \small
	\centering
	\caption{Comparison of prediction performances with different feature sets (``Acc'' is short for accelerometer). }
	\label{tb:prediction_feature}
		\begin{tabular}{cccc}
			\toprule[1.2pt]
			\bf Feature Sets (\#) & \bf Precision & \bf Recall & \bf F1-score   \\ \midrule[0.5pt]
			\bf All (284) & 0.58 & 0.61 & 0.60 \\
			\bf Acc model (78) & 0.44 & 0.63 & 0.52   \\ 
			\bf Phone specifications (3) & 0.48 & 0.56 & 0.51  \\ 
			\bf Location (169) & 0.31 & 0.80 & 0.45 \\
			\bf Acc manufacturer (11) & 0.36 & 0.50 & 0.42   \\ 
			\bf Phone manufacturer (22) & 0.45 & 0.39 & 0.42 \\
			\bf Time (1) & 0.31 & 0.50 & 0.38 \\
	    \midrule
		\end{tabular}
\end{table}




\section{Impact of Accelerometer Data Quality}
\label{impacts}

In this section, we implement quality control measures for real-world earthquake events to evaluate the influence of accelerometer data quality on earthquake parameter estimation. \bigskip





\begin{figure}[tb]
	\centering
	\includegraphics[width=0.43\textwidth]{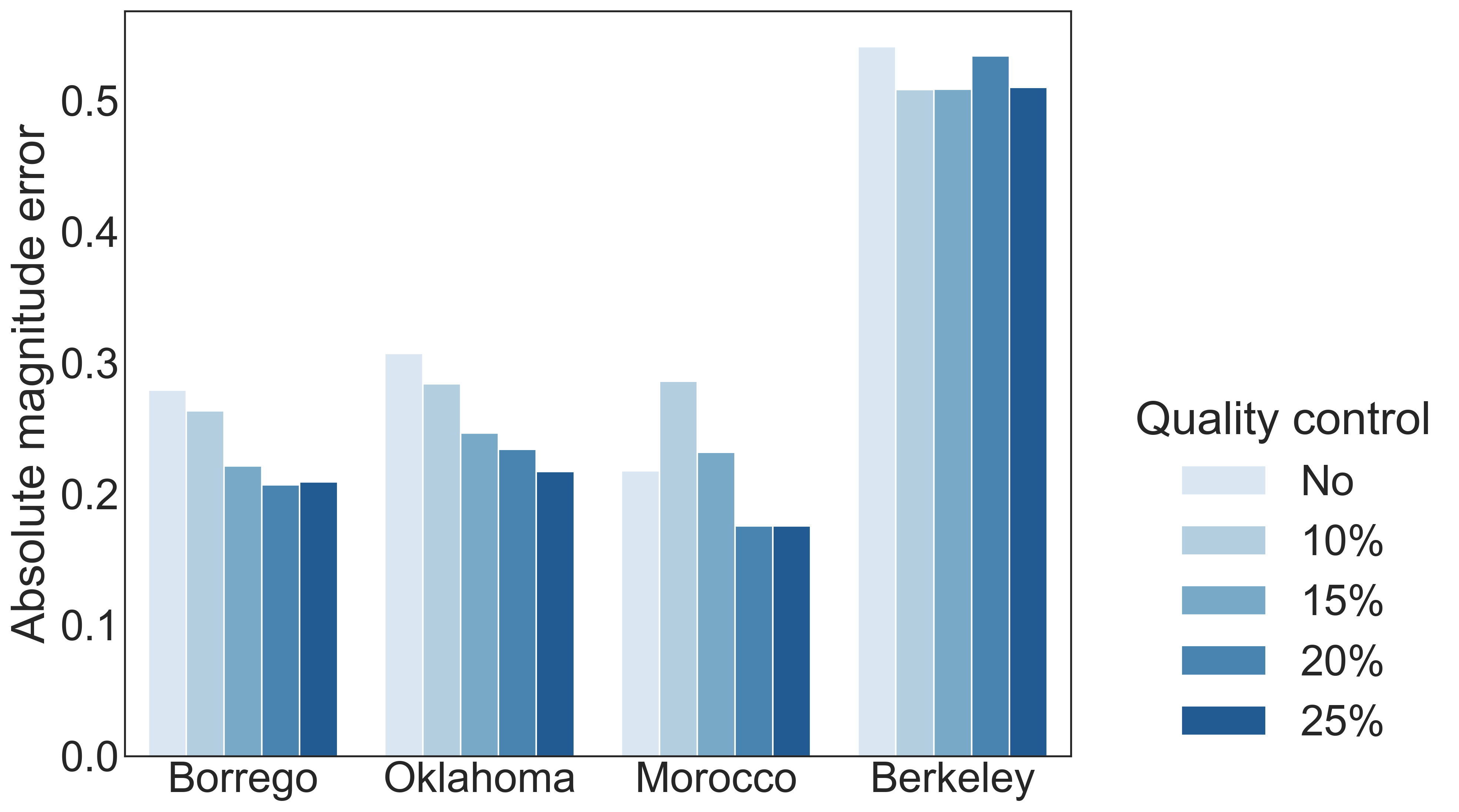}
	\caption{
 Comparison of absolute magnitude estimation errors with different quality control levels versus no quality control. } 
	\label{fig:magnitude_errors}
\end{figure}

To evaluate the impact of accelerometer data quality on earthquake parameter estimation, we examine four earthquake events (outlined in Table~\ref{tb:events}) as case studies. These events vary in magnitude and number of waveforms. We employ the same method used in~\cite{kong2019assessing} to estimate earthquake magnitude, which relies on peak-to-peak amplitude and time span between peaks in seismic waveforms. Fig.~\ref{fig:magnitude_errors} displays the absolute magnitude errors between estimations using different quality control levels and those without. In general, applying quality control to filter out poor-quality waveforms helps reduce the absolute error in magnitude estimation. A 25\% quality control level yields the smallest errors for all four earthquake events. However, the effects of quality control differ among these events. The \textit{Borrego} and \textit{Oklahoma} earthquakes exhibit a similar pattern, with higher quality control levels leading to further error reduction. The \textit{Morocco} earthquake, which has a limited number of waveforms (only 6), shows exceptions with 10\% and 15\% quality control levels. The \textit{Berkeley} earthquake, with its relatively small magnitude, demonstrates less variation in errors across different quality control levels.


\begin{table}[h]
    \small
	\centering
	\caption{Characteristics of the four selected earthquake events. }
	\label{tb:events}
		\begin{tabular}{cccc}
			\toprule[1.2pt]
			\bf Event Name & \bf Time & \bf Magnitude & \bf \# Waveforms   \\ \midrule[0.5pt]
			\bf Borrego & 2016-06-10 & M5.2 & 103 \\ 
			\bf Berkeley & 2018-01-04 & M4.4 & 63 \\
			\bf Oklahoma & 2016-09-03 & M5.8 & 16 \\ 
			\bf Morocco & 2016-03-15 & M5.6 & 6  \\ 
	    \midrule
		\end{tabular}
\end{table}

\section{Conclusions and Future Work}


In this study, we conduct a comprehensive analysis of accelerometer data quality from a global smartphone-based seismic network known as MyShake. We investigate quality issues in the collected waveform data, employing various metrics to assess their sampling rates and noise levels. Additionally, we explore diverse factors, such as phone and accelerometer manufacturer, phone specifications, geolocation, and time, examining their correlation with data quality. Utilizing these factors, we develop quality classification models to identify poor-quality waveforms and assess the importance of various impact factors. Finally, by applying various quality control levels to the collected waveforms, we reveal the influence of data quality on earthquake parameter estimation and present strategies for mitigating these effects. \bigskip

\textbf{Limitations and Future Work.}
In this analysis, we examine approximately four years of accelerometer data gathered by the MyShake system, with waveforms from global devices continuing to accumulate. We focus on key factors influencing accelerometer data quality, recognizing that additional factors and conditions warrant further exploration. Understanding the various impact factors on sensing quality can inform the development of improved strategies to address sensing heterogeneity in real-world applications. Our current quality control analysis investigates only four earthquake events. As future work, we aim to include a broader range of earthquake events and assess the effects of data quality on their key parameter estimation. By considering diverse event characteristics, we can better understand data quality impacts and design more efficient methods to address real-world quality issues. We aspire to enhance the MyShake system by incorporating quality-awareness, ultimately monitoring data quality in real-time and selectively integrating it into the application pipeline. 





\section*{Acknowledgment}
The California Governor’s Office of Emergency Services (Cal OES) funded MyShake through grant 6142-2018 to Berkeley Seismology Lab. Part of this work was performed under the auspices of the U.S. Department of Energy by Lawrence Livermore National Laboratory under Contract Number DE-AC52-07NA27344. Any opinions, findings, conclusions, or recommendations expressed in this publication are those of the authors and do not necessarily reflect those of the supporting agencies. This is LLNL Contribution Number LLNL-JRNL-847732. We thank the global MyShake users and MyShake team at Berkeley for this wonderful project. 

\bibliographystyle{IEEEtran}
\bibliography{main}

\end{document}